
%
%
%
%
\magnification=1200

\hsize=6.5 truein
\vsize=9.0 truein
\parindent=0.5truein
\baselineskip=20pt plus 1.0pt minus 1.0pt
\parskip=10pt plus 0.5pt minus 0.5pt

\font\sixrm=cmr6
\def\ref{\leftskip 30pt\parindent -30pt}
\def\unref{\leftskip 0pt\parindent 30pt}

\def\rmacho {{$R_{\hbox{\sixrm MACHO}}$}}
\def\vmacho {{$V_{\hbox{\sixrm MACHO}}$}}

\def\figlc{{Figure 1}}
\def\figb{{Figure 2}}
\def\figfind{{Figure 3}}
\def\figpower{{Figure 4}}
\def\figplc{{Figure 5}}
\def\figratio{{Figure 6}}

\pageno=2
\headline={\hss\tenrm-- \folio\ --\hss}
\footline={}
\centerline{\bf ABSTRACT}
We report the discovery of 45 beat Cepheids in the Large Magellanic
Cloud (LMC) using the MACHO Project photometry database. The variables
which are pulsating simultaneously in two radial modes are shown
to break cleanly into two period-ratio groups, providing the first
unambiguous evidence that the second overtone is indeed excited in
real Cepheids. Thirty stars are beating in the fundamental and
first overtone mode (F/1H, with a period ratio in the neighborhood of 0.72),
and fifteen stars are beating in the first and second overtone (1H/2H,
with a period ratio near 0.80). The F/1H period ratios are systematically
higher than known Galactic beat Cepheids, indicating a metallicity
dependence whose sense is in agreement with theory. Beat Cepheids in
the LMC are found to select the 1H/2H mode for fundamental periods
shorter than 1.25 days. We find the fraction of Cepheids excited in two
modes to be about 20\% for stars with fundamental periods shorter than
2.5 days. We fail to confirm any of the proposed beat Cepheid
candidates common to our sample from the surveys of Andreasen (1987)
and Andreasen \& Petersen (1987). We also present finder charts and
find several of the beat Cepheids to be in or near LMC clusters.

\vskip 0.2in
{\it Suggested Running head}: LMC Beat Cepheids
\vfil\eject
\centerline{\bf 1. Introduction and Motivation}

Cepheids pulsating in more than one radial mode provide an
excellent test of our understanding of the atmospheres of
supergiant stars. Indeed, to match the observed period ratios
for known beat Cepheids, models employing Cox \& Tabor (1976) opacities
by Petersen (1979) appeared to require masses between 1.0 and 3.0
M$_{\odot}$, far below the 5.0 to 7.0 M$_{\odot}$ predicted by
evolution. This became known as the `Cepheid mass discrepancy', which
was first discussed in detail by Rodgers (1970) using the available
data for the four known galactic `beat Cepheids'.
(A more complete review of the situation at that time is provided
by Cox, 1980). More recently, Petersen (1990) has shown that the use
of the Livermore OPAL opacities (published in Rogers \& Iglesias, 1992)
which are substantially more opaque in the
critical driving region of a Cepheid atmosphere predicts masses
consistent with evolution.

{}From an observational perspective, the situation is not ideal.
When Balona (1985) wrote his review, only 11 galactic double-mode Cepheids
were known. Since that time the star EW Sct has been found to be
a beat Cepheid
by Cuypers (1985) and confirmed by Figer {\it et al.} (1991).
Two additional objects may be added to this list: CO Aur
and AC And. The first was discovered by Mantegazza (1983) and
subsequently characterized by Antonello \& Mantegazza (1984)
and Antonello, Mantegazza, \& Poretti (1986) who claimed that it was
a Cepheid beating in the first and second overtone modes. Balona
(1985) reanalysed their data and disputed this conclusion, claiming
that the peak identified with the second overtone is not distinct
from the noise in the power spectrum. AC And has been identified by
Fitch and Szeidl (1976) as pulsating simultaneously in the fundamental,
first overtone, and second overtone modes. A third principal frequency
was also claimed for TU Cas by Faulkner (1977), the recent history
of the investigation of this variable being described by Matthews {\it
et al.} (1992) who concluded that there is no second overtone present.

The difficulty of establishing the existence (or non-existence) of
second overtone pulsation in Cepheid variables is the result of a
number of circumstances. First, it is very difficult, in practice,
to acquire the long time-series needed to unambiguously identify
power at the appropriate frequencies and their combinations.
This problem is exacerbated by the short periods of many candidates
which are frequently close to multiples of a day. Second,
the tools to remove the undesirable effects of non-uniform sampling
of the lightcurves have only recently become widely available.
Third, candidates are not readily recognized in photographic
variable star surveys (the primary source of variable star discoveries
in the field) as a result of their apparent irregularity and
intervals of low photometric amplitude. Fourth, the lack of convincing
cases argues that only a small percent of beat Cepheids are pulsating in the
second overtone and so a large survey would be required to provide
a meaningful yield. (Only of order 750 galactic Cepheids are known).
Last, even without the preceding difficulties, these stars are
generally found alone and hence independent determinations of their
reddenings and luminosities is not possible. (The sole exception
is the star V367 Sct which lies in the cluster NGC 6649, see
Barrell, 1980). The only surveys undertaken in the Large Magellanic
Cloud (LMC) to date, those of Andreasen (1987) and Andreasen \&
Petersen (1987), were inconclusive.

What questions might profitably be answered by a such a survey?
There are many. Is the second overtone excited? If so, what are
the lightcurve characteristics of this pulsation and is it
observationally distinct from fundamental and first overtone
behavior? What are the actual ratios of periods between the fundamental,
first overtone, and second overtone modes, preferably as a function of
period. Are the ratios a function of metallicity? Do second overtone pulsators
appear in specific regions of the Hertzsprung-Russell diagram? Have
second overtone pulsators been mistaken for first overtone variables
in the past?

To settle these questions, an ideal survey would study a system where
all stars are at a common distance, differential reddening is low,
and large numbers of observations on a common photometric system
are available over a continuous time interval of many hundreds of
cycles. In this paper, we present the results of such a survey. In
section 2, we describe how the observations were obtained. In section 3,
we outline how our sample of beat Cepheid candidates was chosen,
In section 4, we analyse the sample and in section 5 we summarize our
results, describe work in progress, and suggest future avenues of
research.

\bigskip
\centerline{\bf 2. Observations}

The MACHO Project (Alcock {\it et al.} 1992) is an astronomical survey
experiment designed to obtain multiepoch, two-color CCD photometry of
millions of stars in the LMC (also, the galactic bulge and SMC).
The principal goal of the project is to search for massive compact
objects whose presence between the observer and a background source
will result in an amplification of received flux due to gravitational
lensing. The expected rate of detectable events is very low, requiring
large numbers of background sources, in this case, LMC stars, to be
measured over many years.  The survey makes use of a dedicated 1.27m
telescope at Mount Stromlo, Australia and because of its southerly
latitude is able to obtain observations of the LMC year-round. The
camera built specifically for this project (Stubbs {\it et al.} 1993)
has a field-of-view of 0.5 square degrees which is achieved by imaging
at prime focus. Observations are obtained in two bandpasses simultaneously,
using a dichroic beamsplitter to direct the `blue' (approximately 450-630 nm)
and `red' (630-760 nm) light onto 2$\times$2 mosaics of 2048$\times$2048
Loral CCD chips. (Hereafter, these bandpasses will be referred to as
\vmacho\ and \rmacho, respectively). Images are obtained and read out
simultaneously. The 15 $\mu$m pixel size maps to 0.63 arcsec on the sky.
The data were reduced using a profile-fitting photometry routine known
as Sodophot, derived from DoPhot (Mateo \& Schechter, 1989). This
implementation employs a single starlist generated from frames obtained
in good seeing.

The results reported in this survey comprise only a
fraction of the planned data acquisition of the MACHO project.
At present we have processed most of the first year's LMC data,
consisting of some 5500 frames distributed over 22 fields;
this sample contains a total of approximately 8 million stars.
This data has been searched for variable stars and microlensing candidates
and over 40,000 variables have been found, most newly discovered.
The great majority of these fall into four well-known classes: there
are approximately 25000 very red semi-regular or irregular variables,
1500 Cepheids, 8000 RR Lyraes and 1200 eclipsing binaries (Cook, 1995).
Typically the dataset for a given star covers a timespan of about 400 days and
contains 150-320 photometric measurements (multiple observations are
obtained on a given night whenever conditions allow).
Two of our candidate beat Cepheids fall in the overlap region between
two fields and hence have twice the above number of
measurements. The output photometry contains flags indicating suspicion
of errors due to crowding, seeing, array defects, and radiation events.
Only data free from suspected errors was employed for this work,
resulting in output photometry lists of length 100-320. Typical photometric
uncertainties are in the range 1.5-2 percent.

\bigskip
\centerline{\bf 3. Detection of Multimode Cepheids}

Photometry produced by the MACHO Project is currently available in the
form of amplification relative to the median brightness. No transformation
to a standard system was applied prior to our inspection for beat Cepheids.
A star was judged to be variable if it had at
least 7 simultaneous red and blue measurements, with a large $\chi^2$ fit
compared to a constant brightness star in both red and blue,
and a reasonable rank correlation between the red and blue lightcurves.
A total of about 2900 variable stars were identified with median $R_{KC}$ $<$
18.7 and
0.2 $<$ $(V-R)$ $<$ 0.6, where the KC subscript refers to the Kron-Cousins
system. (The transformation to the standard system is only
approximate and is being refined).
The \vmacho\ and \rmacho\ lightcurves were searched for periodicities
using a period-finding code developed by Reimann (1994).  This code
fits either cosines, ``super-smoother'' (Friedman 1984), or periodic cubic
splines to the folded light curve data.  These lightcurves were all fit
using super-smoother.  A weighted sum of absolute residuals was calculated
as a measure of fit which was minimized in a two-step procedure going
from a coarse frequency grid to a fine one.  For these variables, the
best five periods for each light curve were output along with statistics
of the fit.  Since beat
Cepheids are expected to be fit poorly by a single period, about 2900
lightcurves
were examined for evidence of periodic scatter. The Reimann routine
very frequently identified the pattern repetition period as one of the fits
with
the least scatter. A total of 51 stars were judged to have abnormally large
scatter in their lightcurves. Examples of fundamental and overtone
Cepheid \vmacho\ lightcurves are shown in \figlc. Lightcurves for four
LMC beat Cepheids for the longest principal period and the pattern
repetition period are shown in \figb.

Power spectra for the  data were generated using the CLEAN algorithm for
time-series developed by Roberts, Leh\'ar \& Dreher (1987). Briefly, the
raw power spectrum of an irregularly sampled time series is the convolution
of the the actual power spectrum and the spectral window (due to the set
of observing times). The CLEAN algorithm is a way of iteratively removing
the spectral window from the power spectrum to obtain a good approximation
of the actual power spectrum. In our implementation, a gain of 0.5 and 30
iterations were employed. Due to the large number of observations and long
sampling period, the CLEAN algorithm worked very well.

The power spectra of beat Cepheids are distinct from the power spectra of
unresolved double Cepheids. In the latter case, there is significant power
at the two principal frequencies and their multiples. Beat Cepheid
power spectra have significant power at frequencies corresponding to
combinations of the two principal frequencies, as well. A second
discriminant is that double Cepheids appear unusually bright for
the chosen fundamental period in the period-luminosity (P-L) and
period-luminosity-color (P-L-C) diagrams. In fact, their position
reflects the contribution of two sets of fluxes.

In all, we identify 45 LMC beat Cepheid variables. The properties for these
stars are listed in Table 1, where the columns, from left to right are:
right ascension and declination for equinox J2000.0, longer principal
period, shorter principal period, identification of the strongest peak in
the power spectrum as either the longer period (L) or the shorter period (S),
median $R_{KC}$
and $(V-R)$ using the available (and approximate) transformation from
\vmacho\ and \rmacho,
mode identifications where F indicates fundamental mode, 1H indicates
the first overtone, and 2H indicates the second overtone, and comments.
The table is ordered by increasing principal period. (Note that we have
used the term principal period to avoid the initial classification of the
period as fundamental or overtone). The positions have uncertainties less
than 2.0 arcsecs.
Listed separately at the bottom are the three double Cepheid candidates.
The remaining three stars of the original 51 candidates had unusually
large photometric scatter due to other causes or were not Cepheids.
Finder charts for the 48 stars in Table 1 are given in \figfind.

The power spectra for these stars are plotted in \figpower. Note that
asymmetric lightcurves will result in the principal frequency and its
integer multiples being present, whereas a near-sinusoidal shape
will result in most of the power being in the principal peak. The strength
of the frequencies corresponding to mixing modes will be correspondingly
affected.

Of the double-mode candidates found by this survey, only six have previous
identifications in Payne-Gaposchkin (1971). All six have notes indicating
abnormally large scatter in the phased lightcurves. (Cross-identifications
are given in the comments column of Table 1).

We have examined the three LMC double-mode Cepheid candidates proposed by
Andreasen (1987) which are common to our fields. HV 12500, 5694, and
2345 are all normal fundamental mode pulsators with no evidence for a
second (non-commensurate) periodicity. (HV 2345 was considered one of the
two best candidates in that study.) We have also examined the two stars
in common with the list of beat Cepheid candidates given by Andraesen
and Petersen (1987). Neither HV 5664 nor HV 913 show evidence for a second
periodicity.

\bigskip
\centerline{\bf 4. Analysis}

\bigskip
\leftline{\bf a) Strategy}

The fact of the stars being at a common distance (50 kpc) is an enormous
advantage in the interpretation of the observed properties.
The total absorption at $R_{KC}$ is known to be of order 0.15 mags for
field Cepheids and the differential reddening is also correspondingly
low. Hence, the interpretation of observed brightnesses is relatively
straightforward. Before describing our strategy in more detail, it is
useful to briefly review the observational properties of Cepheids.

Cepheid variables are radial pulsators which are found in a regime of
luminosity and effective temperature popularly known as the Cepheid
Instability Strip (CIS). This strip has a finite width in effective temperature
which results in a finite width in observed color. Lines of constant
period and constant luminosity are not parallel and hence a sample
of Cepheid with identical periods can display a range of luminosities.
Since the luminosity is a function of two variables (radius and effective
temperature), we expect that the luminosity of a Cepheid will be better
predicted by the use of a color and a period. Indeed this is found
to be the case --- the P-L-C relation is generally a much tighter fit
than the P-L relation. As is well known, the change in color due to a
difference in effective temperature and due to differential absorption
by interstellar dust grains is very similar. Hence, if we form a
projection of the P-L-C relation which removes the effects of differing
effective temperature, we automatically remove most of the effects of
differential absorption.

In this paper, we are most concerned with mode determination and lightcurve
shape than absolute calibration, so a useful projection of the P-L-C is the
quantity $W_R$, where:
$$W_R\ =\ R\ -\ 4.0(V-R).$$
The exact value of the color coefficient is relatively unimportant, since
as long as it is close to the correct value, the largest part of the small
amount of differential reddening and effective temperature differences will
be removed. The value given was chosen by determining the ratio of
total-to-selective absorption for the $R$ bandpass and the $(V-R)$ color
index, given $A_V$/$E_{B-V}$ = 3.3 (appropriate for the Magellanic
Clouds) using the algorithm of Cardelli, Clayton \& Mathis (1989).
This class of function is described by van den Bergh (1975).

Note that additional light from an unresolved companion will result in
an unusual value of $W_R$. The most common circumstance for contamination
is the presence of an early type (hence bluer) star in the same resolving
element. Since the addition of the light of a blue star will make the
$(V-R)$ color index smaller and increase the brightness at $R$ marginally,
we expect such stars to fall below the main locus of points in the
$W_R$-$\log_{10}$ $P$ diagram. A sensitive test for this contamination
is to determine the ratio of amplitudes between the \vmacho\ and \rmacho\
bandpasses which will be different than for isolated stars.
Note also that a comparison of this type is not possible with galactic
Cepheids since they do not have a common distance.

\bigskip
\leftline{\bf b) Beat Cepheids}

In \figplc, we plot $W_R$ versus $\log_{10}$ $P_L$ for about 1500 stars
identified as Cepheids in our sample.
The periods assumed in this plot are photometric
periods. Also shown are the appropriate values for our beat Cepheid
candidates, assuming the longer of the two principal periods, $P_L$. There
are several feature of this plot which are worthy of comment. First, there are
two obvious loci of stars which are characterized by different lightcurve
shape. The sequence which extends to longer period is due to stars pulsating in
the fundamental mode and these stars typically have skewed lightcurves.
The sequence which appears at brighter $W_R$ is largely due to stars pulsating
in the first overtone. If the correct value of the fundamental period had
been used for these stars, they would also lie along the fundamental
sequence. The separation in $\log_{10}$ $P$ between the two sequences
therefore corresponds to $\log_{10}$ $(P_{1H}/P_{F})$. The plot first
presented by Cook (1995) shows these same features, but is an \rmacho\
P-L relation.

The beat Cepheids plotted in \figplc\ also fall along two different sequences.
Since we have already used the power spectra of these stars to select the
longer of the two principal periods, this indicates that the beat Cepheid
points falling on the overtone sequence are pulsating in the first and second
overtones, rather than the fundamental anf first overtone modes. This assertion
is borne out by inspection of the period ratio column in Table 1, where these
stars are {\it all} found to have a higher period ratio than those stars
falling on the fundamental sequence.

Since this database contains variables fainter than \rmacho\ = 18.0 mag, we
expect that the survey is essentially complete throughout the period range
examined. An examination of \figplc\ reveals that among short period Cepheids,
the beat phenomenon is actually quite common. For $P_F$ $<$ 2.5 days,
we detected beat Cepheids among 20\% of the sample. If the beat Cepheid
phase is a transition between the two dominant pulsation modes, then the
fraction of time spent in a transition state is contrained by this number.

In \figratio, we have plotted the ratio of the two principal periods,
$P_S/P_L$ versus $\log_{10}$ $P_L$. Two very tight sequences of points
are seen, dividing very neatly near $\log_{10}$ $P_L$ = 0.1 ($P_L$ = 1.25
days).
For $P_L$ less than 1.25 days, the period ratios are all in the neighborhood
of 0.80, whereas for $P_L$ greater than 1.25 days, the ratios fall near 0.72.
In both cases, there is a dependence of period ratio on $P_L$ in the
sense that the ratio gets larger with decreasing period. Linear fits to the
LMC F/1H and LMC 1H/2H sequences give the following relations:
$$\eqalign{P_S/P_L\ &=\ 0.733\ -\ 0.034\log_{10} P_L,\ \ 0.1 < \log_{10} P_L
< 0.7,\cr
&=\ 0.803\ -\ 0.022\log_{10} P_L,\ \ -0.2 < \log_{10} P_L \le 0.1.\cr}$$
The ratio of the fundamental to {\it second} overtone period can be found
in the neighborhood of $P_L$ = 1.25 days and its value is $P_{2H}/P_{F}$ =
0.585.
Hence, if Cepheids beating in the second overtone alone exist in the LMC
sample,
they will be found displaced in $\log_{10}$ $P_L$ from the fundamental sequence
by $\log_{10}$ 0.585 = -0.233. Since the displacement of the second and first
overtone sequence is only $\log_{10}$ 0.80 = -0.10, any photometric scatter
may result in these two sequences overlapping. Therefore, Cepheids pulsating
solely in the second overtone may exist and may have been confused with
first overtone stars to date. The relatively small scatter in the LMC ratios
is remarkable and must be due to both the uniformity of the observations and
a small spread in metallicity among the LMC sample. The scatter is not
significantly greater than what is expected from the uncertainties in the
periods themselves from the limited timespan of these observations.

Also, shown are data for the 12 galactic beat Cepheids listed by Balona (1985)
plus EW Sct.
There is a clear difference between the period ratios of galactic and LMC
beat Cepheids, presumably due to the difference in metallicity between the
two samples. Luck \& Lambert (1992) found the LMC Cepheids to have [Fe/H] =
-0.3
with a dispersion of 0.3, although this dispersion is inflated by one
very metal-rich 100-day Cepheid in their sample.
(The SMC, by comparison, has [Fe/H] = -0.7.)
The trend in period ratio $P_S/P_L$ for these Cepheids is given by:
$$P_S/P_L\ =\ 0.720\ -\ 0.027\log_{10} P_L,\ \ 0.3 < \log_{10} P_L < 0.8.$$

The sense of the period ratio change with metallicity can be derived from
Petersen (1990) where the ratios predicted using the more recent opacities
of Rogers \& Iglesias (1992) are compared with models using the
Cox \& Tabor (1976)
opacities. The older opacities, because of their lower absorption in the
driving region, can be considered representative of a more metal-poor
atmosphere. Figure 1 in Petersen shows that lower F/1H period ratios are
associated with more opacity in the driving region of the Cepheid atmosphere
and so our observationally derived sense of period-ratio change is in
agreement with theory. More recently Moskalik, Buchler \& Marom (1992)
have calculated models for specific masses and luminosities which show this
same dependence on metallicity. The period ratio for 1H/2H LMC beat Cepheids
is also expected to be higher than for their more metal-rich galactic
counterparts.

\bigskip
\leftline{\bf c) ``Double Cepheids''}

In our sample of beat Cepheid candidates, we have identified three stars
whose photometric scatter appears to be due to the superposition of the
light variations of two Cepheids. The two characteristics which result
in this classification are the absence of power at frequencies corresponding
to the mixing modes of the two principal frequencies and brightnesses
which are clearly the sum of two normal Cepheids. The last three panels
of \figpower\ contain the power spectra for these stars. Let us discuss these
in turn.

\noindent{\bf MACHO*05:21:54.8-69:21:50} The lightcurve of this system
is due to two 1H pulsators as evidenced by the weakness of the power
spectra peaks at twice the principal frequencies (and lack of peaks that
exceed the noise at higher multiples). There is no evidence for any excess
power at the difference and sum frequencies. The 0.55 mag displacement
brightward of the 1H sequence in \figplc\ is consistent with the sum
of the light from two 1H pulsators of photometric period 2.48 and 1.96
days, which would result in a brightness difference of 0.61 mags.

\noindent{\bf MACHO*04:59:17.5-69:14:18} The power spectrum of this
object is clearly the result of the light variations of a F and a 1H
pulsator. The principal frequency and its second, third, and fourth
multiples are seen in the power spectra, indicating a skew lightcurve.
This is consistent with its identification as a fundamental pulsator.
No higher multiples of the second principal frequency are seen, so the
the second star is clearly a 1H pulsator. No excess signal is seen at the
sum and difference frequencies of the two principal frequencies.
The 0.59 mag displacement brightward of the F sequence in \figplc\
is consistent with these identifications. Using a period ratio of
0.715 for the 1H star, we predict a brightness increase of 0.72 mags
for stars of the same color.

\noindent{\bf MACHO*05:04:02.3-68:21:32} The power spectrum for this
system reveals it to consist of two F pulsators. The second, third,
and fourth multiples of both of the two principal frequencies are
evident and no mixing modes are seen. This object lies 0.41 mags
above the ridgeline of fundamental sequence in \figplc. This is
consistent with the prediction of an increase of 0.47 mags for two
fundamental pulsators with the given periods.

The question of whether or not these stars are physically related cannot
be answered with these observations. It is clear that they must be very
similar in age and mass, or they would not be sufficiently similar in
period and luminosity to be detected in this search. A radial velocity
study of these systems could provide evidence for orbital motion. Given
the spectroscopic binary mass function and the assumption that the two
stars have equal mass, it would be possible to derive lower limits for
the masses of the stars. However, the shortest period Cepheid binaries
in our galaxy have orbital periods of a year and a half, so such a
programme would require long timescale monitoring.

There is one example of a double Cepheid in our galaxy: CE a+b Cas.
Sandage and Tammann (1969) first discussed these two variables, which
are located in the galactic cluster NGC 129. They are separated by
about 8000 AU, so no detectable orbital motion is expected or seen.
At the distance of the LMC, this pair of objects would be unresolved
by this survey.

\bigskip
\leftline{\bf c) Beat Cepheids Near LMC Clusters}

Our understanding of the evolution of stars and the evolutionary
context of variable star phenomena has benefited from the discovery
of variable stars in clusters. The Magellanic Cloud clusters are
especially useful because there are many rich, young clusters which
have few or no counterparts in the observable volume of our Galaxy.
In preparing finder charts for the 48 stars in Table 1, it became
clear that several of the beat Cepheids are found at positions
very close to both known and uncatalogued clusters. We will discuss
the associations we believe to be most promising.

\noindent{\bf MACHO*04:54:55.0-69:14:12}
This star is found adjacent to the cluster NGC 1756. There are
three additional Cepheids associated with this
cluster. All of them are F pulsators with periods between 2.67
and 3.5 days.

\noindent{\bf MACHO*05:36:54.7-70:08:10}
This variable is found near both NGC 2059 and NGC 2058, the latter
being located just south of the boundary of the finder chart. NGC 2059
appears to be misidentified on chart 53B of Hodge \& Wright (1967).
There are five Cepheids associated with this relatively sparse cluster.
Four are 1H variables with periods between 2.13 and 3.31 days, and one
is a fundamental pulsator with a period of 5.58 days.
There are at least 6 additional Cepheids associated with NGC 2058.
Three of these are F pulsators with periods between
4.68 and 5.35 days, and three are 1H pulsators with periods between
1.92 and 2.09 days. Additional variables found in the immediate
neighborhood may also belong. Bica, Clari\'a, \& Dottori (1992)
derive an equivalent cluster type of III on the Searle, Wilkinson \&
Bagnuolo (1980) classification scheme (SWB type). Class II and III
clusters are where Cepheid variables are most likely to be found.
Using the calibration of Elson \& Fall (1985), NGC 2058 has an age
of approximately 110 million years.

\noindent{\bf MACHO*05:33:39.4-69:54:55}
This variable is found to be in the midst of the sparse cluster
HS 353 (from the cluster catalogue of Hodge \& Sexton, 1966).

\bigskip
\centerline{\bf 5. Conclusions and Future Work}

The principal conclusions of this work are:
1) we have identified 45 beat Cepheids in the LMC, thirty of which
are F/1H pulsators,
2) fifteen of these stars are found to have a period ratio near
0.80 and a brightness which indictates that they are pulsating
in the 1H/2H modes,
3) the F/1H period ratios are systematically higher in the LMC
sample than in the galactic sample, indicating a structural
difference between LMC and galactic Cepheids,
4) Cepheids in the LMC select the 1H/2H modes for periods
shorter than 1.25 days,
5) among stars with fundamental periods shorter than 2.5 days,
double-mode excitation is seen about 20\% of the time,
6) the sense of the period-ratio/metallicity dependence
agrees with that expected from pulsation theory,
7) we fail to confirm any of the beat Cepheid candidates
proposed to date (which are common to our dataset), and
8) we find several beat Cepheids to be in or near LMC clusters, some
of which contain other Cepheid variables.

There are many avenues of Cepheid research which may yet be
explored with the MACHO photometry database. We will describe
a few here.

The final analysis of some properties of the beat Cepheids reported
in this paper must await the final photometric transformations
between the instrumental MACHO photometry and a standard system.
This analysis is underway and should be avilable soon.

There are almost certainly stars pulsating solely in the
second overtone which have not been recognized thus far due
to the overlap of the 1H and 2H sequences in the P-L relation.
We have seen evidence for systematic differences in lightcurve
shape and are actively pursuing the characterization of the three
modes. Stellingwerf, Gautschy \& Dickens (1987) have predicted
that 2H lightcurves will be asymmetric like F pulsators but
distinct.

The relative properties of SMC beat Cepheids (if they are found
to exist) will be especially interesting in view of the much lower
metallicity of this sample. Differences in the regions of the
color-magnitude diagram occupied by Cepheids in different modes
will provide strong constraints for pulsation models. The SMC
fields which exist in the MACHO photometry database have not yet
been searched for variables.

The difference in period ratios is evidence that LMC Cepheids
metallicity affects the atmospheres of Cepheids in an observable
way. With this knowledge in hand, we will explore the effects
on the P-L and P-L-C relations.

The full MACHO dataset will span at least four years. During this
time a sensitive test for time-variable mode strengths may be
undertaken. This is the only dataset which is sufficiently long,
contains sufficient numbers of stars and is of sufficient quality
to test for such changes.

We are very grateful for the skilled support given our project
by the technical staff at the Mt. Stromlo Observatory.
Work performed at LLNL is supported by the DOE under contract W7405-ENG-48.
Work performed by the Center for Particle Astrophysics on the UC campuses
is supported in part by the Office of Science and Technology Centers of
NSF under cooperative agreement AST-8809616.
Work performed at MSSSO is supported by the Bilateral Science
and Technology Program of the Australian Department of Industry, Technology
and Regional Development. KG acknowledges a DOE OJI grant, and CWS
and KG thank the Sloan Foundation for their support.
DLW was a Natural Sciences and Engineering Research Council (NSERC)
University Research Fellow during this work.
This research has made use of data obtained from the
Canadian Astronomy Data Centre, which is operated by the Dominion
Astrophysical Observatory for the National Research Council of Canada's
Herzberg Institute of Astrophysics, as well as the NASA's Astrophysics
Data System (ADS) Abstract Service.

\centerline{\bf References}
\ref

Alcock, C. {\it et al.} 1992, in ``Robotic Telescopes in the 1990s'',
ASP Conf. Ser. No. 34, ed. A.V. Fillippenko, p.193

Andreasen, G.K. 1987, A\&A, 186, 159

Andreasen, G.K., and Petersen, J.O. 1987, A\&A, 180, 129

Antonello, E., and Mantegazza, L. 1984, A\&A, 133, 52

Antonello, E., Mantegazza, L., and Poretti, E. 1986, A\&A, 159, 269

Balona, L.A. 1985, in ``Cepheids: Theory and Observations'', IAU
Colloquium No. 82, ed. B.F. Madore (Cambridge: Cambridge University
Press), p.17

Barrell, S.L. 1980, ApJ, 240, 145

Bica, E., Clari\'a, J.J., and Dottori, H. 1992, AJ, 103, 1859

Cardelli, J.A., Clayton, G.C., and Mathis, J.S. 1989, ApJ, 345, 245

Cook, K.H. 1995, in ``Stellar Populations'', IAU Symposium No. 167,
(Dordrecht: D. Reidel)

Cox, A.N. 1980, ARAA, 18, 15

Cox, A.N. and Tabor, J.E. 1976, ApJS, 31, 271

Cuypers, J. 1985, A\&A, 145, 283

Faulkner, D.J. 1977, ApJ, 218, 209

Figer, A., Poretti, E., Sterken, C., and Walker, N. 1991,
MNRAS, 249, 563

Friedman, J.H. 1984, ``A Variable Span Smoother'', Technical Report
No. 5, Laboratory for Computational Statistics, Department of Statistics,
Stanford University

Hodge, P.W., and Sexton, J.A. 1966, AJ, 71, 363

Hodge, P.W., and Wright, F.W. 1967, ``The Large Magellanic Cloud'',
(Smithsonian Press, Washington, D.C.)

Luck, R.E., and Lambert, D.L. 1992, ApJS, 79, 303

Mantegazza, L. 1983, A\&A, 118, 321

Mateo, M., and Schechter, P. 1989, Proceedings of the 1st ESO-ECF Data
Analysis Workshop, edited by P.J. Grosb{\o}l, F. Murtagh, and R.H. Warmels,
(European Southern Observatory, Garching)

Matthews, J.M., Gieren, W.P., Fernie, J.D., and Dinshaw, N. 1992,
AJ, 104, 748

Moskalik, P., Buchler, J.R., and Marom, A. 1992, ApJ, 385, 685

Payne-Gaposchkin, C.H. 1971, ``The Variable Stars of the Large Magellanic
Cloud'', Smithsonian Contributions to Astrophysics, 13, 1

Petersen, J.O. 1990, A\&A, 238, 160

Petersen, J.O. 1979, A\&A, 80, 53

Reimann, J. 1994, Ph.D. thesis, University of California at Berkeley,
Department of Statistics, ``Frequency Estimation Using Unequally-Spaced
Astronomical Data''

Roberts, D.H., Leh\`ar, J., and Dreher, J.W. 1987, AJ, 93, 968

Rodgers, A.W. 1970, MNRAS, 151, 133

Rogers, F.J., and Iglesias, C.A. 1992, ApJS, 79, 507

Sandage, A., and Tammann, G.A. 1969, ApJ, 157, 683

Searle, L., Wilkinson, A., and Bagnuolo, W. 1980, ApJ, 239, 803

Stellingwerf, R.F., Gautschy, A., and Dickens, R.J. 1987, ApJ, 313, L75

Stubbs, C.W. {\it et al.} 1993, in ``Charge-coupled Devices and Solid
State Optical Sensors III'', ed. M. Blouke, Proc. of the SPIE, 1900, 192

van den Bergh, S. 1975, ``The Extragalactic Distance Scale,'' in
{\it Stars and Stellar Systems}, Volume 5: Galaxies and the Universe,
eds. A. Sandage, M. Sandage, and J. Kristian, (Chicago: University of
Chicago Press), p.509

Welch, D.L., Mateo, M., and Olszewski, E.W. 1993, in ``New Perspectives
on Stellar Pulsation and Pulsating Variable Stars'', IAU Colloquium No. 139,
eds. J.M. Nemec and J.M. Matthews, (Cambridge: Cambridge University Press),
p.359


\unref
\vfil\eject
\centerline{\bf Figure Captions}

\bigskip\bigskip
\figlc --- Lightcurves (\vmacho) are shown for four fundamental mode (F)
LMC Cepheids (left panels) and four first overtone (1H) Cepheids (right
panels). The major tick marks on the ordinate are 0.5 mag apart. Each point
is plotted twice (separated in phase by 1.0) to preserve continuity. Error
bars are plotted. The star designation containing equinox J2000.0 coordinates
is given on the upper left of each panel and the period used (to 0.01 days)
is given in the upper right.

\bigskip\bigskip
\figb --- Lightcurves (\vmacho) are shown for four LMC beat Cepheids.
The left panel contains the lightcurve phased with period $P_L$ from
Table 1. The right panel is the same photometry phase with the
pattern-repetition period $P$ = $(2P_LP_S)/(P_L-P_S)$. The period
used for plotting is not precisely that found from using the values
from Table 1, but the period corresponding to the minimum `string
length' in the neighborhood of the expected pattern-repetition period.

\bigskip\bigskip
\figfind --- Finder charts for the variables listed in Table 1. The size of
the image portion of each chart is 120 $\times$ 120 arcsec. The label appears
on the south side of the image, and west is to the right. The coordinates
are the equinox J2000.0 positions for the variables - not the chart center.
In cases where the star was near the edge of the image, the region was
selected to show as much field as possible and consequently the star is not
always at the chart center. These are \rmacho\ images and have been stretched
to show fainter stars. Known cluster identifications are shown. However,
it is obvious that many of the beat Cepheids are found in or projected on
uncatalogued clusters of stars. Both NGC 1976 and NGC 2058 contain other
Cepheid variables.

\bigskip\bigskip
\figpower a-f --- Power spectra for beat Cepheid candidates identified in this
survey are shown. The ordinate is the logarithm of the power
at that frequency. Based on the positions of the two principal frequencies,
the mixing frequencies, as well as multiples of the principal frequencies,
are identified with arrows. The symbols \rlap{$\bigcirc$}L, \rlap{$\bigcirc$}S,
$\ominus$, $\oplus$, \rlap{$\bigcirc$}{\sixrm L2}, \rlap{$\bigcirc$}{\sixrm
S2},
represent the longer and shorter principal frequencies,their difference and
sum,
and their second multiples, respectively. The star designation is also given.
Note that if the shape of the lightcurve in one of the modes is nearly
sinusoidal
the higher multiples of the corresponding principal frequency (and their
mixing modes) will be effectively absent. The last three panels contain
power spectra identified as being due to ``double Cepheids'' -- two Cepheids
that appear within the same resolving element on the sky (and possibly
related).
Note the absence of mixing mode frequencies in the power spectra for these
stars.

\bigskip\bigskip
\figplc --- The quantity $W_R$, which removes the greatest fraction of
brightness difference due to differential reddening and color difference,
is plotted against the logarithm of the photometric period. Two sequences
of Cepheids are seen in this diagram. For a given period, the lower sequence
is due to stars pulsating in the fundamental (F) mode. The upper sequence is
primarily due to stars pulsating in the first overtone (1H). This sequence
appears
brighter because the photometric period is shorter than the fundamental period
appropriate for these stars. Beat Cepheids pulsating in the F/1H mode are shown
as
solid circles and stars pulsating in the 1H/2H modes are shown as open circles.
Data for these stars are found in Table 1. The circles containing `B'
correspond to
``double Cepheids'' --- two Cepheids which are seen in the same resolving
element. Stars falling in the lower right-hand part if the diagram are Type II
Cepheids and RV Tau stars for the most part.

\bigskip\bigskip
\figratio --- The ratio of the shorter photometric period, $P_S$, to the longer
photometric period, $P_L$, is plotted as a function of the logarithm of $P_L$
for LMC and galactic beat Cepheids. The stars with $P_S$/$P_L$ near 0.72 are
identified as beating in the fundamental and first overtone (F/1H) modes,
whereas the stars near $P_S$/$P_L$ = 0.80 are identified as pulsating in the
first
and second overtone (1H/2H) modes. Filled circles and open circles correspond
to F/1H and 1H/2H LMC beat Cepheids, respectively. Open squares indicate
galactic beat Cepheids --- all F/1H pulsators.  Near $P_L$ = 1.25 days, the
ratio
of the second overtone to fundamental period is derived to be 0.585. The period
ratios for LMC Cepheids are systematically higher than for galactic Cepheids,
presumably due to differences in metallicity. The lines shown are unweighted
linear least-squares fits to the appropriate data. The equations for these
lines are given in the text.
\bigskip\bigskip
\vfil\eject\bye